\documentclass[10pt,a4paper]{article}

\usepackage{amsfonts}
\usepackage{amssymb}
\usepackage{amsmath}
\usepackage{latexsym}

\usepackage{epsfig}
\usepackage{graphicx}
\usepackage{bm}

\title{\bf \Large  Entropy of nonlinear black holes\\
in quadratic gravity}

\author{Jerzy Matyjasek\footnote{jurek@kft.umcs.lublin.pl, matyjase@tytan.umcs.lublin.pl}
\\
\noalign{\vspace{3ex}}
\it{ \small Institute of Physics,  Maria Curie-Sk\l odowska University,}\\
\it{\small pl. Marii Curie-Sk\l odowskiej 1, 20-031 Lublin, Poland}}

\date{}

\begin{document}

\maketitle
\begin{abstract}
Employing the Noether charge technique and  Visser's Euclidean approach
the entropy of the nonlinear black hole described by the perturbative
solution of the system of coupled equations of the quadratic gravity and
nonlinear electrodynamics is constructed. The solution is parametrized by the
exact location of the event horizon and charge. Special emphasis in put on
the extremal configuration. Consequences of the second choice of the
boundary conditions, in which the solution is paramerized by the charge and
the total mass as seen by a distant observer is briefly examined.
\end{abstract}

\section{\protect\bigskip Introduction}

Recently, a great deal of efforts have been devoted to the important
issue of regular black holes. One of the most intriguing solutions of
this type have been constructed by Ay\'{o}n-Beato and Garc\'{i}a
\cite{ABG} and by Bronnikov \cite{Bronnikov1}. In both cases, the line
element is a solution of the coupled system of equations of nonlinear
electrodynamics and gravity. (We shall refer to the solutions of this
type as ABGB geometries). The former solution describes a regular,
static and spherically symmetric configuration with the electric
charge, $Q_{e},$ whereas the latter one describes a similar geometry
characterized by the mass and the magnetic charge $Q$. For certain
values of the parameters both solutions describe black holes. On the
other hand, the no-go theorem proved in Ref.~\cite{Bronnikov3} (see
also~\cite{Bronnikov1,Bronnikov2}) forbids, for the class of
electromagnetic Lagrangians with a Maxwell asymptotic in a weak field
limit, existence of the electrically charged, static and spherically-
symmetric solutions with the regular center. It should be noted,
however, that the electric solution is not in conflict with the non
existence theorem, as the formulation of the nonlinear
electrodynamics~\cite{Plebanski} employed by Ay\'{o}n-Beato and
Garc\'{i}a (P framework in the nomenclature of Ref. \cite{Bronnikov1}
) differs from the one to which one refers in the assumptions of the
no-go theorem. Indeed, the solution of Ay\'{o}n-Beato and Garc\'{i}a
has been constructed in a formulation of the nonlinear electrodynamics
obtained from the original one (F framework) by means of a Legendre
transformation (see Ref. \cite{Bronnikov1} for details). Moreover,
the no-go theorem does not forbid existence of the solutions with
magnetic charge as well as some hybrid configurations in which the
electric field does not extend to the central region.

The status of the nonlinear electrodynamics in the model considered
here is to provide a static matter source, perhaps the exotic one, to
the field equations. That means that  the casual structure of the
spacetime is still governed by the null geodesics or ``ordinary"
photons rather than the photons of the nonlinear theory. Actually, the
latter move along the geodesics of the effective
space~\cite{Novello1,Novello2}. Outside the event horizon the solution
of the ABGB-type closely resembles the Reissner-Nordstr\"om (RN)
geometry both in its global and local structure. Important differences
appear near the extremality limit. Consequently, the Penrose diagrams
of the ABGB solution are similar to those constructed for the Rissner-
Nordstr\"om solution,  with the one notable distinction: instead of
the singularity at $r=0$ now we have the regular interior.

An attractive feature of the ABGB solutions is possibility to express
the location of the horizons in terms of the Lambert special functions
\cite {Kocio1,Kocio2a}. Similarly, the Lambert
functions~\cite{Lam1,Lam2} may be used in the discussion of the
extremal configurations \cite{Kocio3}.


According to our present understanding a proper description of the
gravitational phenomena should be given by the quantum gravity, being
perhaps a part of a more fundamental theory. And although at the
present stage we have no clear idea how this theory looks like, we
expect that the action functional describing its low-energy
approximation should consist of the higher order terms constructed
from the curvature tensor, its contractions and covariant derivatives
to some required order. Among various generalizations of the Einstein-
Hilbert action a special role is played by the quadratic gravity (see
for example Refs.~\cite{Utiyama,Stelle1,Stelle2,Weyl1,Weyl2,Pauli,
sirArthur,Steve,Whitt:1984pd}). Motivations for introducing such terms
into the action functional are numerous. When invented, for example,
the equations of quadratic gravity have been treated as an exact
formulation of the theory of gravitation. On the other hand, it may be
considered, quite naturally, as truncation of series expansion of the
action of the more general theory. Such terms appear generically in
the one-loop calculations of the quantum field theory in curved
background \cite{Birrell}. Moreover, from the point of view of the
semi-classical gravity, the quadratic terms in the field equations
might be treated as some sort of the simplified stress-energy tensor.
Such a toy model of the renormalized stress- energy tensor allows to
mimic the fairly more complex sources in a relatively simple way. This
approach is especially useful when the general pattern that lies
behind the calculations of both types is essentially the same. Thus,
some general features of the full semi- classical solutions can be
analyzed and understand without referring to otherwise intractable
equations.

It should be noted that any higher curvature theory contain solutions
which are unavailable to the theory based on the classical Einstein-
Hilbert Lagrangian. This can most easily be seen by counting the
degrees of freedom: the quadratic gravity is known to posses 8 degrees
of freedom whereas the General Relativity has only 2. Moreover, there
are solutions that are not analytic in the coupling constants, i. e.,
they do not reduce to solutions of the classical Einstein field
equations. (For a comprehensive discussion see for example \cite{LS} 
and the references cited therein).
Unfortunately, because of  complexity of the equations of the
quadratic gravity it is practically impossible to construct their
exact solutions and one is forced to refer either to approximations or
to numerical methods. The natural method to obtain reasonable results
consists of treating the higher curvature contributions
perturbatively. This approach also guarantees that the black hole
exists as the perturbative solution of the higher-order solution
provided it exists classically~\cite{Myers1}. Finally, observe that in
the perturbative approach the casual structure is determined by the
classical metric, however, the equations of motion of test particles
and various characteristics of the solution acquire the first order
correction.

Analyses of the spherically-symmetric and static solutions to the
higher derivative theory has been carried out in \cite
{Stelle1,Lousto1,Lousto2,Lousto3,Holdom1,Tryn,ja_grg}. Specifically,
in Ref. \cite{ja_grg} the perturbative solutions of the ABGB-type to
the equations of the effective quadratic gravity  have been
constructed and discussed. In this paper we shall calculate the
entropy of such black holes using Wald's approach
\cite{Wald1,Iyer,Jac1} and confirm the final results employing
computationally independent but closely related Euclidean techniques
propounded by Visser \cite{Visser1,Visser2,Visser3}.

\section{\protect\bigskip Basic equations}

The coupled system of the nonlinear electrodynamics and the quadratic
gravity considered in this paper is described by the (Lorentzian) action 
\begin{equation}
S=\frac{1}{16\pi}\int\left( R+\alpha R^{2}+\beta R_{ab}R^{ab}+\gamma
R_{abcd}R^{abcd}-\mathcal{L}\left( F\right) \right) \sqrt{-g}\,d^{4}x, 
\end{equation}
where $\mathcal{L}\left( F\right) $ is some functional of $F=F_{ab}
F^{ab}$ (its exact form will be given later) and all symbols have
their usual meaning. The cosmological constant is assumed to be zero.
To simplify our discussion from the very beginning we shall relegate
the term involving the Kretschmann scalar, $R_{abcd}R^{abcd}$, from
the total action employing the Gauss-Bonnet invariant.  The coupling
constants $\alpha$ and $\beta$ have the dimension of length squared
and throughout the paper we shall assume
\begin{equation}
\frac{\alpha}{L^{2}} \sim \frac{\beta}{L^{2}} \ll 1,
                     \label{small}
\end{equation}
where $L$ is the local curvature scale. Assumption that the mass
scales associated with the linearized equations are real may place
additional constrains~\cite {Steve,Whitt:1985,Audretsch:1993kp} on
$\alpha$ and $\beta.$ Here, however, we shall treat them as small and
of comparable order but arbitrary.

The entropy of the black hole may be calculated using various methods.
It seems, however, that Wald's technique is especially well suited for
calculations in the higher curvature theories. Here we shall follow
this very approach. Other competing techniques are the method based on
the field redefinition \cite{Jac1,Jac2} and Visser's Euclidean
approach.

For the Lagrangian involving the Riemann tensor and its symmetric
derivatives up some finite order, say $n,$  Wald's Noether charge
entropy may be compactly written in the form \cite{Wald1,Iyer,Jac1}
\begin{equation}
\mathcal{S}=-2\pi \int d^{2}x\left( h\right) ^{1/2}\sum_{m=0}^{n}\left(
-1\right) ^{m}\nabla _{(e_{1}...}\nabla
_{e_{m})}Z^{e_{1}...e_{m};abcd}\epsilon _{ab}\epsilon _{cd}, 
                                                \label{wentr}
\end{equation}
where 
\begin{equation}
Z^{e_{1}...e_{m};abcd}=\frac{\partial \mathcal{L}}{\partial \nabla
_{(e_{1}...}\nabla _{e_{m})}R_{abcd}},
\end{equation}
$h$ is the determinant of the induced metric, $\epsilon _{ab}$ is the
binormal to the bifurcation sphere, and the integration is carried out
across the bifurcation surface. Actually $\mathcal{S}$ can be evaluated not
only on the bifurcation surface but on an arbitrary cross-section of the
Killing horizon. Since $\epsilon _{ab}\epsilon _{cd}=\hat{g}_{ad}\hat{g}
_{bc}-\hat{g}_{ac}\hat{g}_{bd}$, where $\hat{g}_{ac}$ is the metric in the
subspace normal to cross section on which the entropy is calculated, one can
rewrite Eq. (\ref{wentr}) in the form 
\begin{equation}
\mathcal{S}=4\pi \int d^{2}x\,h^{1/2}\sum_{m=0}^{n}\left( -1\right)
^{m}\nabla _{(e_{1}...}\nabla _{e_{m})}Z^{e_{1}...e_{m};abcd}\hat{g}_{ac}
\hat{g}_{bd}.  \label{wald1}
\end{equation}
The tensor $\hat{g}_{ab}$ is related to $V^{a}=K^{a}/||K||$ ($K^{a}$ is the
timelike Killing vector) and the unit normal $n^{a}$ by the formula $\hat{g}
_{ab}=V_{a}V_{b}+n_{a}n_{b}.$

The general expression describing entropy (\ref{wald1}) has been applied in
numerous cases, mostly for the Lagrangians that are independent of covariant
derivatives of the Riemann tensor and its contractions. In Ref. \cite
{kocio_2006_3}, however, Eq. (\ref{wald1}) has been employed in calculations
of the entropy of the quantum-corrected black hole when the source term is
described by the stress-energy tensor of the quantized fields in a large
mass limit. Such a tensor is purely geometrical and besides ordinary higher
curvature terms it involves also $R\nabla _{a}\nabla ^{a}R$ and $
R_{ab}\nabla _{c}\nabla ^{c}R^{ab}.$

On the other hand, one can follow an approach propounded by Visser \cite
{Visser1,Visser2,Visser3}. The general formula for the entropy of the
stationary black hole with the Hawking temperature $T_{H}$ \ is given by 
\begin{equation}
\mathcal{S=}\frac{A}{4}+\frac{1}{T_{H}}\int_{\Sigma }\left( \rho
_{L}-L_{E}\right) K^{a}d\Sigma _{a}+\int_{\Sigma }\mathbf{s}V^{a}d\Sigma
_{a},  \label{visser}
\end{equation}
where $A$ is the area of the event horizon, $\mathbf{s}$ is the entropy
density associated with the fluctuations (ignored in this paper) and finally 
$\rho _{L}$ and $L_{E}$ are, respectively, the Lorentzian energy density and
the Euclideanized Lagrangian of the matter fields surrounding the black
hole. (All higher curvature terms have been inserted into the Lagrangian
describing matter fields.) For the specific case of the Einstein-Hilbert
action augmented with the higher curvature terms (but not covariant
derivatives of curvature) Visser's result is equivalent to Wald's formula.

The coupled system of differential equations describing nonlinear
electrodynamics in quadratic gravity can be obtained from the variational
principle. 

Simple calculations indicate that the tensor $F^{ab}$ and its dual
$^{\ast}F^{ab},$ satisfy the equations
\begin{equation}
\nabla_{a}\left( \dfrac{d\mathcal{L}\left( F\right) }{dF}F^{ab}\right) =0,
\end{equation}
\begin{equation}
\nabla_{a}\,^{\ast}F^{ab}=0,
\end{equation}
respectively. Differentiating functionally the total action $S$ with
respect to the metric tensor one obtains equations of the quadratic
gravity in the form
\begin{equation}
L^{ab}\equiv G^{ab}-\alpha I^{ab}-\beta J^{ab}=8\pi T^{ab},
\label{2nd_order}
\end{equation}
where 
\begin{equation}
I^{ab}=2\nabla^{b}\nabla^{a}R-2RR^{ab}+\frac{1}{2}g^{ab}\left(
R^{2}-4\nabla_{c}\nabla^{c}R\right) ,
\end{equation}
\begin{equation}
J^{ab}=\nabla^{b}\nabla^{a}R-\nabla_{c}\nabla^{c}R^{ab}-2R_{cd}R^{cbda}+
\frac{1}{2}g^{ab}\left( R_{cd}R^{cd}-\nabla_{c}\nabla^{c}R\right)
\end{equation}
and 
\begin{equation}
T_{a}^{b}=\dfrac{1}{4\pi}\left( \dfrac{d\mathcal{L}\left( F\right) }{dF}
F_{ca}F^{cb}-\dfrac{1}{4}\delta_{a}^{b}\mathcal{L}\left( F\right) \right) .
\label{ep}
\end{equation}

In this paper we shall concentrate on the static and spherically-symmetric
configurations described by the line element of the form 
\begin{equation}
ds^{2}=-e^{2\psi\left( r\right) }f(r)dt^{2}+\frac{dr^{2}}{f(r)}
+r^{2}d\Omega^{2},  \label{el_gen}
\end{equation}
where 
\begin{equation}
f(r)=1-\frac{2M(r)}{r}.  \label{with_f}
\end{equation}
The spherical symmetry places restrictions on the components of $F_{ab}$
tensor and, consequently, its only nonvanishing components compatible with
the assumed symmetry are $F_{01}$ and $F_{23}$. Simple calculations yield 
\begin{equation}
F_{23}=Q\sin\theta
\end{equation}
and 
\begin{equation}
r^{2}e^{-2\psi}\dfrac{d\mathcal{L}\left( F\right) }{dF}F_{10}=Q_{e},
\end{equation}
where $Q$ and $Q_{e}$ are the integration constants interpreted as the
magnetic and electric charge, respectively. 

Since the no-go theorem forbids existence of the regular solutions
with $Q_{e} \neq 0$ in the latter we shall assume that the electric
charge vanishes. Now, since $F=2F_{23}F^{23},$ one has
\begin{equation}
F=\dfrac{2Q^{2}}{r^{4}}.  \label{postacF}
\end{equation}
The stress-energy tensor (\ref{ep}) calculated \ for this configuration is 
\begin{equation}
T_{t}^{t}=T_{r}^{r}=-\dfrac{1}{16\pi}\mathcal{L}\left( F\right)  \label{t1}
\end{equation}
and 
\begin{equation}
T_{\theta}^{\theta}=T_{\phi}^{\phi}=\dfrac{1}{4\pi}\dfrac{d\mathcal{L}\left(
F\right) }{dF}\dfrac{Q^{2}}{r^{4}}-\dfrac{1}{16\pi}\mathcal{L}\left(
F\right) .  \label{t2}
\end{equation}

Further considerations require specification of the Lagrangian $\mathcal{L}
\left( F\right) .$ Following Ay\'{o}n-Beato, Garc\'{i}a and Bronnikov let us
chose it in the form 
\begin{equation}
\mathcal{L}\left( F\right) \,=F\left[ 1-\tanh ^{2}\left( s\,\sqrt[4]{\frac{
Q^{2}F}{2}}\right) \right] ,  \label{Lagr}
\end{equation}
where 
\begin{equation}
s=\frac{\left| Q\right| }{2b},  \label{ss}
\end{equation}
and $b$ is a free parameter. Inserting Eq.~(\ref{postacF}) into (\ref{Lagr})
and making use of Eq.~(\ref{ss}) one obtains 
\begin{equation}
\mathcal{L}\left( F\right) =\frac{2Q^{2}}{r^{4}}\left( 1-\tanh ^{2}\frac{
Q^{2}}{2br}\right) .  \label{LodF}
\end{equation}

The system of coupled differential equations of the quadratic gravity
with the source term given by (\ref{t1}) and (\ref{t2}) with
(\ref{LodF}) is rather complicated and cannot be solved exactly.
Fortunately, since the coupling constants $\alpha $ and $\beta $ are
expected to be small in a sense of Eq.~\ref{small}, one can treat the
system of the differential equations perturbatively, with the
classical solution of the Einstein field equation taken as the zeroth-
order approximation. Successive perturbations are therefore solutions
of the chain of the differential equations of ascending
complexity~\cite{Simon1,Simon:1991jn,Parker1,SO}. It should be noted,
however, that the higher order equations are probably intractable
analytically and the technical difficulties may limit the calculations
to the first order.

In the next section, we shall employ perturbative techniques to
construct the approximate solution to the equations of the quadratic
gravity with the source term being the stress-energy tensor of the
Bronnikov type. Such an approach is expected to yield reasonable
results and because of complexity of the differential equations, it
may be the only way to deal with this problem.

\section{Solutions}

To keep control of the order of terms in complicated series expansions
we shall introduce a dimensionless parameter $ \varepsilon$
substituting $\alpha\rightarrow\varepsilon\alpha$ and 
$\beta\rightarrow\varepsilon\beta$. We shall put $\varepsilon=1$ at the
final stage of calculations. Of functions $M\left( r\right) $ and
$\psi\left( r\right) $ we assume that they can be expanded in powers
of the auxiliary parameter as
\begin{equation}
M\left( r\right) =M_{0}\left( r\right) +\varepsilon M_{1}\left( r\right) +
\mathcal{O}\left( \varepsilon^{2}\right)  \label{Mser}
\end{equation}
and 
\begin{equation}
\psi\left( r\right) =\varepsilon\psi_{1}\left( r\right) +\mathcal{O}\left(
\varepsilon^{2}\right) .  \label{psiser}
\end{equation}

First, consider the left hand side of Eq. (\ref{2nd_order}) calculated
for the line element (\ref{el_gen}) with the functions $M(r)$ and
$\psi(r)$ given by (\ref{Mser}) and (\ref{psiser}), respectively.
Making use of the above expansions and subsequently collecting the
terms with the like powers of $\varepsilon,$ after some
rearrangements, one obtains~\cite{ja_grg}
\begin{equation}
L_{t}^{t}=-\frac{2}{r^{2}}(M_{0}^{\prime}+\varepsilon M_{1}^{\prime
}-\varepsilon S_{t}^{t}),  \label{1st}
\end{equation}
where 
\begin{align}
S_{t}^{t} & =\beta\left( \frac{2\,M_{0}^{\prime}}{r^{2}}-\frac
{8\,M_{0}\,M_{0}^{\prime}}{r^{3}}+\frac{2\,{M_{0}^{\prime}}^{2}}{r^{2}}-
\frac{2\,M_{0}^{\prime\prime}}{r}+\frac{5\,M_{0}\,M_{0}^{\prime\prime}}{r^{2}
}-\frac{M_{0}^{\prime}\,M_{0}^{\prime\prime}}{r}\right.  \notag \\
& \left. +\frac{{M_{0}^{\prime\prime}}^{2}}{2}+M_{0}^{(3)}-\frac
{M_{0}\,M_{0}^{(3)}}{r}-M_{0}^{\prime}\,M_{0}^{(3)}+r\,M_{0}^{(4)}-2\,M_{0}
\,M_{0}^{(4)}\right)  \notag \\
& -\alpha\left( \frac{24\,M_{0}\,M_{0}^{\prime}}{r^{3}}-\frac{
8\,M_{0}^{\prime}}{r^{2}}-\frac{4\,{M_{0}^{\prime}}^{2}}{r^{2}}+\frac{
8\,M_{0}^{\prime\prime}}{r}-\frac{18\,M_{0}\,M_{0}^{\prime\prime}}{r^{2}}-{
M_{0}^{\prime\prime}}^{2}\right.  \notag \\
& \left. +\frac{2\,M_{0}^{\prime}\,M_{0}^{\prime\prime}}{r}-4\,M_{0}^{(3)}+
\frac{6\,M_{0}\,M_{0}^{(3)}}{r}+2\,M_{0}^{\prime}\,M_{0}^{(3)}-2\,r
\,M_{0}^{(4)}+4\,M_{0}\,M_{0}^{(4)}\right)  \notag \\
&  \label{1sta}
\end{align}
and $M_{0}^{\prime},$ $M_{0}^{\prime\prime}$ and $M_{0}^{(i)}$ for $i\geq3$
denote first, second and $i-$th derivatives with respect to the radial
coordinate. On the other hand, a simple combination of the components of $
L_{a}^{b}$ tensor 
\begin{equation}
L_{r}^{r}-L_{t}^{t}=0  \label{2nd}
\end{equation}
can be easily integrated to yield~\cite{ja_grg} 
\begin{equation}
\psi_{1}(r)\,=\,(2\alpha+\beta)M_{0}^{(3)}-{\frac{4}{r^{2}}}(3\alpha
+\beta)M_{0}^{\prime}+C_{1},  \label{pseq}
\end{equation}
where $C_{1}$ is the integration constant. It should be noted that
contrary to the case of coupled system of the Maxwell equations and
quadratic gravity considered in
Refs.~\cite{Lousto1,Lousto2,Lousto3,Tryn}, now we have explicit
dependence on the parameter $\alpha.$ A comment is in order here
regarding the independence of the final result calculated for the
Maxwell source on the parameter $\alpha.$ First, observe that the
stress-energy tensor of the electromagnetic field for the spherically-
symmetric an static configuration with a total charge $e$ assumes
simple form
\begin{equation}
T_{a}^{b} = - \frac{e^2}{8 \pi r^{4}} {\rm diag}[1,1,-1,-1].
\end{equation}
Therefore, the zeroth-order solution to the (0,0) -component 
of the equation (\ref{2nd_order}) can be written in the form
\begin{equation}
M_{0}(r) = -\frac{e^2}{2 r}+ C,
                            \label{rr_nn}
\end{equation}
where $C$ is the integration constant. Now, substituting (\ref{rr_nn})
into (\ref{1sta}) and (\ref{pseq}) it can easily be demonstrated that
the expression in the second bracket in its right hand side of
Eq.~(\ref{1sta})  as well as the expression $M_{0}^{3}-6 M'_{0}/r^{2}$
in (\ref{pseq}) vanish.

One expects that all characteristics of the black hole, such as the
location of the horizons and temperature could also be calculated
perturbatively. In the latter, for simplicity, we shall refer to the
perturbative solutions of the quadratic gravity using the names of
their classical counterparts (the zeroth-order solutions) whenever it
will not lead to confusion.

To develop the model further one has to determine the integration
constants and the free parameter $b$. There are, in general, two
interesting and physically motivated choices. One can relate the
integration constant with the exact location of the event horizon,
$r_{+}$, and this can easily be done with the aid of the equation
\begin{equation}
M\left( r_{+}\right) =\frac{r_{+}}{2}.  \label{bound0}
\end{equation}
On the other hand it is possible to express solutions of the system of
differential equations consisting of $\left( 0,0\right) $ component of
Eqs. ( \ref{2nd_order}) and Eq. (\ref{2nd}) in terms of the total mass
$\mathcal{M}$ as seen by a distant observer
\begin{equation}
\lim_{r\rightarrow\infty}M\left( r\right) =\mathcal{M.}  \label{bound1}
\end{equation}
For the function $\psi\left( r\right) $ we shall always adopt the natural
condition 
\begin{equation}
\lim_{r\rightarrow\infty}\psi\left( r\right) =0.  \label{bound2}
\end{equation}

Inspection of Eqs. (\ref{1st}) and (\ref{2nd}) reveals their different
status. Indeed, Eq. (\ref{2nd}) can easily be integrated for a general
function $M_{0}(r)$ and the final solutions is to be obtained by
differentiation of the zeroth-order solution and making use of the
boundary conditions. On the other hand, the first integral of the
differential equation for $M_{1}\left( r\right) $ cannot be
constructed and one has to know the zeroth-order solution to determine
$M_{1}.$

The assumed expansions of the functions $M\left( r\right) $ and
$\psi\left( r\right) $ as given by Eqs. (\ref{Mser}) and
(\ref{psiser}), respectively, suggests that one can rewrite the
boundary conditions of the first type in the following form:
\begin{equation}
M_{0}\left( r_{+}\right) =\frac{r_{+}}{2},\quad M_{1}\left( r_{+}\right)
=0,\quad \psi _{1}\left( \infty \right) =0,  \label{1st_type}
\end{equation}
whereas for the boundary conditions of the second type one has 
\begin{equation}
M_{0}\left( \infty \right) =\mathcal{M,\quad }M_{1}\left( \infty \right)
=0,\quad \psi _{1}\left( \infty \right) =0.  \label{2nd_type}
\end{equation}


Now, let us concentrate on the zeroth-order equations supplemented with 
the conditions of the first type. Putting $\varepsilon =0$ in Eq.~(\ref{1st}),
form (\ref{LodF}) and (\ref{t1}) one obtains
\begin{equation}
\frac{d M_{0}}{dr} = \frac{Q^{2}}{2 r^{2}}\left(1 -\tanh^{2} 
\frac{Q^{2}}{2 b r}\right),
\end{equation}
which can be easily integrated to yied
\begin{equation}
M_{0}(r) = -b\tanh\frac{Q^{2}}{2br} + C_{2}.
\end{equation}
Finally, making use of the conditions (\ref{1st_type}) one arrives at the desired 
result 
\begin{equation}
M_{0}\left( r\right) =\frac{r_{+}}{2}+b\tanh \frac{Q^{2}}{2br_{+}}-b\tanh 
\frac{Q^{2}}{2br}.  \label{mm}
\end{equation}
The thus obtained solution reduces to the Schwarzschild solution for $Q=0$
and it can be easily demonstrated that, by (\ref{1sta})and the boundary 
conditions (\ref{1st_type}) it remains so in the higher-order
calculations.

To specify the solution further we shall make use of the
well-known relation \cite{Hawking_Bardeen} 
\begin{equation}
\mathcal{M}=\frac{\kappa A_{H}}{4\pi }-\int_{\Sigma }\left(
2T_{a}^{b}-T\delta _{a}^{b}\right) K^{a}d\Sigma _{b},  \label{hawking}
\end{equation}
where $\Sigma $ is a constant time hypersurface and $K^{a}$ is a timelike
Killing vector and apply it to the zeroth-order solution. Making use of the
explicit form of the stress-energy tensor of the nonlinear electrodynamics
one obtains

\begin{equation}
M_{H}=\frac{r_{+}}{2}+b\tanh \frac{Q^{2}}{2br_{+}},
\end{equation}
where $M_{H}$ is the mass connected with the zeroth-order solution. We shall
refer to $M_{H}$ as to the horizon defined mass of the black hole.

To develop the model further one has to determine the free parameter $b.$
Our choice, which guarantees regularity of the zeroth-order line element
at the center, is $b=M_{H},$ and hence Eq. (\ref{mm}) becomes 
\begin{equation}
M_{0}\left( r\right) =M_{H}\left( 1-\tanh \frac{Q^{2}}{2M_{H}r}\right) .
\label{mmm}
\end{equation}
Unfortunately, the regularity of the zeroth-order solution does not
guarantee regularity of the higher-order perturbative solutions \cite{ja_grg}.

It should be noted that the $M_{H}=M_{H}\left( Q,r_{+}\right) $ and  for
fixed $Q$ and $r_{+}$ one has to determine $M_{H}$ numerically. On the other
hand it is possible to employ the equation $M\left( r_{+}\right) =r_{+}/2$
in the zeroth-order calculations and express all the results in $\left(
Q,M_{H}\right) $ parametrization instead of $(Q,r_{+}).$ One can, therefore,
construct solutions of this equation in terms of the Lambert function.
Simple manipulations yield 
\begin{equation}
r_{+}=-\frac{4M_{H}Q^{2}}{4W_{+}\left( -\rho e^{\rho }\right) M_{H}^{2}-Q^{2}
},  \label{hor}
\end{equation}
where $W_{+}$ is a principal branch of the Lambert function and $\rho
=Q^{2}/4M_{H}^{2}.$ Analogous solution for the inner horizon can be written
in the form 
\begin{equation}
r_{-}=-\frac{4M_{H}Q^{2}}{4W_{-}\left( -\rho e^{\rho }\right) M_{H}^{2}-Q^{2}
},
\end{equation}
where $W_{-\text{ }}$ is the second real branch of the Lambert function. (In
fact, $W_{+}$ and $W_{-}$ are the only real branches.)  
Making use of the elementary properties of the Lambert functions one 
can demonstrate that the principal branch has the expansion
\begin{equation}
W_{+}(x) = x -x^{2} + \frac{3}{2}x^{3} - \frac{8}{3} x^{4} +{\cal O}(x^{5}). 
\end{equation} 
On the other hand,  $W_{-}(x) \to -\infty$ as  $x \to 0,$ and,
consquently, the location of the event horizon tends to the Schwarzschild value 
whereas $r_{-} \to 0.$

A typical run of $M_{H}$ as a function of $\xi$ for a few exemplary values
of $Q$ is showed in Fig. 1. For a given $Q$  a line of $M_{H}=const.$
intersects $Q = const.$ curve at one or two points or it has no intersection points at
all. The smaller one gives location of the inner horizon whereas the greater
is to be identified with the event horizon. The minimum of 
$M_{H} =M_{H}\left(Q=const,\xi\right)$ function represents extremal configurations 
when the two horizons merge. 
\begin{figure}[h]
\includegraphics[height=6cm]{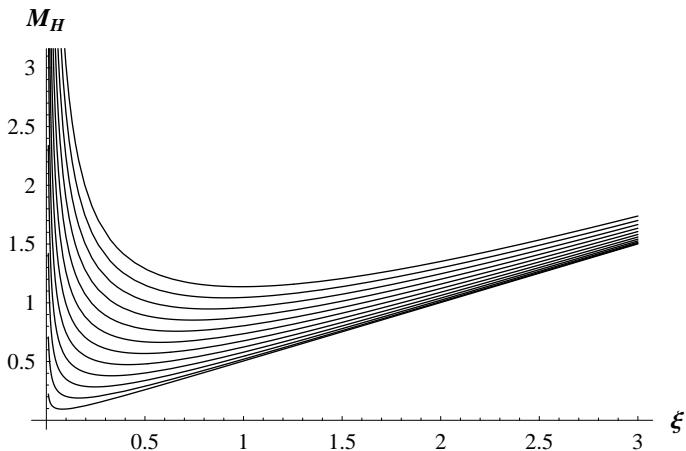}
\caption{This graph shows solutions of the equation 
$M_{H} =\frac{\xi}{2} +M_{H} \tanh\frac{Q^2}{2 M_{H}\xi}$ 
for a few exemplary values of the charge $Q.$
From bottom to top the curves correspond to $Q=0.1i,$ for $i=1,...,12.$ For $
M_{H}=const.,$ the greater solution represents location of the event horizon, $r_{+}$
whereas the smaller one represents the inner horizon, $r_{-}$. The minimum of each
curve corresponds to the extremal configuration with $r_{+}=r_{-}.$}
\label{f1}
\end{figure}

It should be noted that the mass $M_{H}$ is not the mass that would be
measured for the perturbed black hole by an observer at infinity. Indeed,
even for the zeroth-order solutions the meaning of $M_{H}$ and $\mathcal{M}$ 
is different and the substantial differences 
are transparent in the first order calculations. This can be easily seen 
by studying the limit 
\begin{equation}
\mathcal{M}=\lim_{r\rightarrow \infty }M\left( r\right) =M_{H}+\varepsilon
M_{1}\left( \infty \right)  \label{limit}
\end{equation}
and Eq. (\ref{bound1}). Identical result can be obtained form Eq.
(\ref {hawking}). Indeed, in order to apply (\ref{hawking}) for the
perturbed black hole one has to move the higher curvature terms of Eq.
(\ref{2nd_order} ) into its right hand side and treat them as a
contribution to the total stress-energy tensor. It can be demonstrated
explicitly, that making use of Eq. (\ref{hawking}) in the first-order
calculations one obtains precisely (\ref{limit}).

The function $M_{1}(r)$ can be expressed in terms of the
polylogarithms. Unfortunately, it is rather complicated and to avoid
unnecessary proliferation of long formulas it will not be displayed
here. The first-order solution can be constructed employing the
algorithm presented in Appendix of Ref.~\cite{ja_grg}. It should be
noted that the function $M_{1}(r)$ presented in~\cite{ja_grg} is
calculated for the boundary conditions of the second type.

\section{The entropy}

Now, let us return to our main theme and calculate the entropy of the ABGB
black hole. In doing so we shall put special emphasis on comparison of the
results constructed for the nonlinear black hole with the analogous results
obtained for the Reissner-Nordstr\"om solution. Such a comparison is
especially interesting as the geometries of their classical counterparts are
practically indistinguishable in two important regimes. To demonstrate this
it suffices to expand the metric potentials in powers of $|Q|/r_{+}$ and $
r_{+}/r,$ respectively.
Since the expansion takes the form
\begin{equation}
f(r)\,=\, 1- \frac{2M_{H}}{r} + \frac{Q^{2}}{r^{2}}-\frac{Q^{6}}{12 M_{H}^{2}r^{4}} + ...,
\end{equation} the differences in the metric structure between 
ABGB and RN geometries in the exterior region for $|Q|/r_{+}\ll 1$ are small indeed. 
One has a similar behaviour for any (allowable) value of the charge for $r\gg r_{+}$.

The higher curvature terms in the action functional lead to the appearance
of additional terms in the final expression describing entropy, which spoil
area/entropy relation. Simple calculations carried out within the Noether
charge framework indicate that the contribution of the quadratic part of the
action to the entropy is given by 
\begin{equation}
\delta \mathcal{S}=2\pi r_{+}^{2}\left[ \alpha R+\frac{1}{2}\beta \left(
R_{t}^{t}+R_{r}^{r}\right) \right] _{|r_{+}}.  \label{S2}
\end{equation}
Now, substituting the line element (\ref{el_gen}) with (\ref{with_f}) and (
\ref{Mser}), into the general expression (\ref{S2}), expanding the right
hand side of Eqs. (\ref{2nd_order}) with respect to $\varepsilon ,$ and,
finally, retaining the linear terms only, one gets 
\begin{equation}
\mathcal{S}=\pi r_{+}^{2}+2\pi \varepsilon r_{+}^{2}\left[ \frac{4\alpha }{
r_{+}^{2}}M_{0}^{\prime }\left( r_{+}\right) +\frac{2\alpha +\beta }{r_{+}}
M_{0}^{\prime \prime }\left( r_{+}\right) \right] +\mathcal{O}\left(
\varepsilon ^{2}\right) .  \label{S_gen}
\end{equation}
For the nonextreme black hole with the boundary conditions of the first type
(\ref{1st_type}) one has 
\begin{align}
\mathcal{S}=&\pi r_{+}^{2}+\frac{2\pi Q^{2}}{M_{H}r_{+}^{3}}\varepsilon \cosh
^{-2}\left( \frac{Q^{2}}{2M_{H}r_{+}}\right) \left\{ \alpha Q^{2}\tanh
\left( \frac{Q^{2}}{2M_{H}r_{+}}\right) \right.\notag \\
&\left.
-\beta \left[ M_{H}r_{+}-\frac{Q^{2}
}{2}\tanh \left( \frac{Q^{2}}{2M_{H}r_{+}}\right) \right] \right\} ,
\label{een}
\end{align}
where $M_{H}=M_{H}\left( Q,r_{+}\right) $. In this approach the zeroth-order
solution (\ref{mmm}) determines the first order correction to the entropy of
the nonextreme black hole completely. Having established $M_{H}$ for given $
Q $ and $r_{+}$ one can rewrite Eq. (\ref{een}) putting $\mathcal{\tilde{S}}=
\mathcal{S}/M_{H}^{2}$, $q=\left| Q\right| /M_{H}$, $x_{+}=r_{+}/M_{H\text{ }
}$, $\tilde{\alpha}=\alpha /M_{H}^{2}$ and $\tilde{\beta}=\beta /M_{H}^{2}.$
Simple manipulations yield 
\begin{equation}
\mathcal{\tilde{S}}=\pi x_{+}^{2}+\varepsilon \frac{2\pi q^{2}}{x_{+}^{3}}
\cosh ^{-2}\frac{q^{2}}{2x_{+}}\left[ \tilde{\alpha}Q^{2}\tanh \frac{q^{2}}{
2x_{+}}-\tilde{\beta}\left( x_{+}-\frac{q^{2}}{2}\tanh \frac{q^{2}}{2x_{+}}
\right) \right] .  \label{ABGB_ent}
\end{equation}
This results can be contrasted with the analogous result constructed for the
Reissner-Nordstr\"om black hole

\begin{equation}
\mathcal{S}=\pi r_{+}^{2}-2\beta \frac{\pi Q^{2}}{r_{+}^{2}}  \label{en_rn1}
\end{equation}
or 
\begin{equation}
\mathcal{\tilde{S}}=\pi (1+\sqrt{1-q^{2}})^{2}-2\tilde{\beta}\frac{\pi q^{2}
}{(1+\sqrt{1-q^{2}})^{2}}\text{.}  \label{en_rn2}
\end{equation}
To investigate the entropy $\mathcal{S}$ as given by Eq. (\ref{een}) let us
observe that for $\left| Q\right| /r_{+}\ll 1$ one has $r_{+}\approx 2M_{H}.$
Now, expanding hyperbolic functions in powers of $\left| Q\right| /r_{+}$
one obtains 
\begin{equation}
\mathcal{S}=\pi r_{+}^{2}-2\pi \beta \frac{Q^{2}}{r_{+}^{2}}+\mathcal{O}
\left( (\frac{Q}{r_{+}})^{4}\right) .  \label{en_abg_small}
\end{equation}
A comparison of Eqs. (\ref{en_abg_small}) and (\ref{en_rn1}) shows that for $
\left| Q\right| /r_{+}\ll 1$ the entropies of the ABGB and RN black holes are
almost indistinguishable, as expected. It should be noted that contrary to
the Reissner-Norstr\"om geometry, the entropy of the ABGB black hole depends
on $\alpha $ and for $\left| Q\right| /r_{+}\ll 1$ the leading behaviour of
this terms is $\sim \left( Q/r_{+}\right) ^{4}.$

The analysis of the extremal configuration is more involved. First, let us
return to the zeroth-order solution. It should be emphasized that although
we do not ascribe any particular meaning to the zeroth-order solution, some
of its features do possess clear and unambiguous meaning. For the boundary
conditions (\ref{1st_type}) such a solution is described by the exact $r_{+}$
and $Q.$ The extremality condition places additional relation between the
elements of the pair $(Q,r_{+})$ or $\left( Q,\mathcal{M}\right) $. Here we
shall confine ourselves to the first pair. Simple considerations yield 
\begin{equation}
\left| Q\right| =2w^{1/2}M_{H}  \label{relat1}
\end{equation}
and 
\begin{equation}
r_{+}=\frac{4w}{1+w}M_{H},  \label{relat2}
\end{equation}
where $w=W_{+}(1/e)$, and consequently 
\begin{equation}
\left| Q\right| /r_{+}=\frac{1+w}{2w^{1/2}}.  \label{relat3}
\end{equation}
Returning to the first-order solution we recall the relation valid for the
extremal configuration in the Reissner-Nordstr\"om geometry

\begin{equation}
r_{+}=\left| Q\right| .
\end{equation}
In Ref.~\cite{Kocio3} we have ascribed this simple relation to
tracelessness of the stress-energy tensor of the matter fields. As the
stress-energy tensor of the nonlinear electrodynamics considered in this
paper has a nonzero trace, one expects that the analogous relation between $Q
$ and $r_{+}$ in the ABGB geometry is more complicated. Indeed, after some
algebra, one has 
\begin{equation}
r_{+}=\frac{2w^{1/2}\left| Q\right| }{\left( 1+w\right) }\left[
1+\varepsilon \frac{\beta +2\alpha }{16Q^{2}w}\left( w+3\right) \left(
w^{2}-1\right) \right] .  \label{rp_vs_Q}
\end{equation}
Now, making use of (\ref{rp_vs_Q}) in (\ref{een}) gives 
\begin{equation}
\mathcal{S}_{extr}=\frac{4\pi wQ^{2}}{\left( 1+w\right) ^{2}}-\frac{\pi
\varepsilon }{2\left( 1+w\right) }\left[ \left( 2\alpha +\beta \right)
w^{2}+2\left( 2\alpha -\beta \right) w+5\alpha +2\beta \right],
\label{exttr}
\end{equation}
and the first term of the right hand side coincides with the Bekenstein-
Hawking entropy~\cite{Myung}.
Numerically, one has 
\begin{equation}
\mathcal{S}_{extr}=\pi Q^{2}\times 0.6815-2\pi \varepsilon \times \left(
0.0324\alpha +0.1754\beta \right), 
\end{equation}
where a common factor $2\pi $ has been singled out for convenience.
Analogous relation for the extremal Reissner-Nordstr\"om black hole reads 
\begin{equation}
\mathcal{S}_{extr}=\pi Q^{2}-2\pi \varepsilon \beta .
\end{equation}

Now, let us calculate the entropy of the ABGB black hole employing the
Euclidean techniques propounded by Visser. First, observe that if the
Lagrangian is arbitrary function of the Riemann tensor (and its
contractions) but is independent of its covariant derivatives, both methods,
i. e. Wald's approach and Visser's method are equivalent. One may wonder,
therefore, why we intend to carry out such a calculation. The answer is
simple: although both methods should yield identical results, the
calculational steps necessary to obtain the final result are quite different
and consequently one can consider the calculations carried out within the
framework of the one method as the useful check of the other. It is
especially important in situations when the computational complexity of the
considered problem may lead to numerous errors.

The calculations proceed in a few steps. First, incorporate the Euclidean
action functional of the quadratic gravity into the matter part of the
action. Similarly, the (Lorentzian) energy density is given by 
\begin{equation}
\rho=-T_{t}^{t}=\frac{1}{16\pi}\mathcal{L}\left( F\right) -\varepsilon
\left( \frac{\alpha}{8\pi}I_{t}^{t}+\frac{\beta}{8\pi}J_{t}^{t}\right) .
\label{roro}
\end{equation}
It could easily be demonstrated that $\rho_{L}-L_{E}\sim O\left(
\varepsilon\right) $ and consequently it suffices to know the Hawking
temperature to the zeroth-order. Moreover, due to subtle cancellations in
the integrand of Eq. (\ref{visser}) the final result of the quadratures does
not contain polylogarithm functions. Now, substituting 
\begin{equation}
T_{H}=\frac{1}{4\pi r_{+}}\left( 1-\frac{Q^{2}}{Mr_{+}}+\frac{Q^{2}}{
4M_{H}^{2}}\right)  \label{th}
\end{equation}
and (\ref{roro}) into (\ref{visser}), after some algebra, one has
\begin{equation}
\delta\mathcal{S}=\frac{\varepsilon}{r_{+}^{4}\left( \eta+1\right) ^{5}}
\left[ \alpha s_{\alpha}+\beta s_{\beta}\right] ,  \label{entr_viss1}
\end{equation}
where $\eta=\exp\left( Q^{2}/2M_{H}r_{+}\right) ,$

\begin{align}
&s_{\alpha } =\left( \,{\frac{4{Q}^{6}}{\mathit{r_{+}}\,{\mathit{M_{H}^{2}}}}
}-\,{\frac{20{Q}^{4}}{\mathit{M_{H}}}}\right) {\eta }^{4}-\left( {\frac{20{Q}
^{4}}{\mathit{M_{H}}}}-{\frac{72{Q}^{4}}{\mathit{r_{+}}}}+\,{\frac{12{Q}^{6}
}{\mathit{r_{+}}\,{\mathit{M_{H}^{2}}}}}+{\frac{8{Q}^{6}}{{\mathit{r_{+}^{2}}
}\mathit{M_{H}}}}\right) {\eta }^{3}+  \notag \\
& \left( {\frac{20{Q}^{4}}{\mathit{M_{H}}}}-{\frac{12{Q}^{6}}{\mathit{r_{+}}
\,{\mathit{M_{H}^{2}}}}}+\,{\frac{56{Q}^{6}}{{\mathit{r_{+}^{2}}}\mathit{
M_{H}}}}\right) {\eta }^{2}+\left( {\frac{4{Q}^{6}}{\mathit{r_{+}}\,{\mathit{
M_{H}^{2}}}}}-\,{\frac{72{Q}^{4}}{\mathit{r_{+}}}}+\,{\frac{20{Q}^{4}}{
\mathit{M_{H}}}}-\,{\frac{16{Q}^{6}}{{\mathit{r_{+}^{2}}}\mathit{M_{H}}}}
\right) \eta  \label{entr_viss2}
\end{align}
and 
\begin{align}
s_{\beta }& =\left( 8\,{Q}^{2}\mathit{r_{+}}+{\frac{2{Q}^{6}}{\mathit{r_{+}}
\,{\mathit{M_{H}^{2}}}}}-\,{\frac{12{Q}^{4}}{\mathit{M_{H}}}}\right) {\eta }
^{4}  \notag \\
& -\left( 24\,\mathit{M_{H}}\,{Q}^{2}+\,{\frac{4{Q}^{6}}{{\mathit{r_{+}^{2}}}
\mathit{M_{H}}}}+{\frac{12{Q}^{4}}{\mathit{M_{H}}}}-24\,{Q}^{2}\mathit{r_{+}}
\text{+}{\frac{6{Q}^{6}}{\mathit{r_{+}}\,{\mathit{M_{H}^{2}}}}}-{\frac{36{Q}
^{4}}{\mathit{r_{+}}}}\right) {\eta }^{3}  \notag \\
& +\left( {\frac{12{Q}^{4}}{\mathit{M_{H}}}}-{\frac{6{Q}^{6}}{\mathit{r_{+}}
\,{\mathit{M_{H}^{2}}}}}-{\frac{8{Q}^{4}}{\mathit{r_{+}}}}+{Q}^{2}\mathit{
r_{+}}+\,{\frac{28{Q}^{6}}{{\mathit{r_{+}^{2}}}\mathit{M_{H}}}}-48\,\mathit{
M_{H}}\,{Q}^{2}\right) {\eta }^{2}  \notag \\
& -\left( 24\,\mathit{M_{H}}\,{Q}^{2}-8\,{Q}^{2}\mathit{r_{+}}\text{ }-{
\frac{12{Q}^{4}}{\mathit{M_{H}}}}-{\frac{2{Q}^{6}}{\mathit{r_{+}}\,{\mathit{
M_{H}^{2}}}}}+{\frac{44{Q}^{4}}{\mathit{r_{+}}}}+\,{\frac{8{Q}^{6}}{{\mathit{
r_{+}^{2}}}\mathit{M_{H}}}}\right) \eta  \label{entr_viss3}
\end{align}
At first glance this result does not resemble Eq. (\ref{een}). However,
making use of the identity 
\begin{equation}
\eta =\frac{4M_{H}}{r_{+}} - 1,
\end{equation}
one can easily demonstrate that Eqs. (\ref{entr_viss1}-\ref{entr_viss3})
reduce precisely to Eq. (\ref{een}).

\section{Final remarks}

In this paper we have constructed the entropy of the nonlinear ABGB-type
black holes using the boundary conditions (\ref{1st_type}).  The
zeroth-order solution coincides, as expected, with the ABGB line element
whereas the first-order correction can be elegantly expressed in terms of
the polylogarithm functions. Now, let us briefly discuss the consequences of
the second choice, in which the results are expressed in terms of the total
mass of the system as measured by a distant observer. To calculate the
location of the event horizon to the required order in $\left( Q,\mathcal{M}
\right) $ parametrization one has to solve the first-order equations for $
M_{1}\left( r\right) $ and $\psi _{1}\left( r\right) ,$ and, subsequently,
perturbatively solve the equation $g_{tt}(r_{+})=0$ assuming that the event
horizon can be expanded as 
\begin{equation}
r_{+}=r_{0}+\varepsilon r_{1}+O(\varepsilon ^{2}).  \label{REH}
\end{equation}
Unfortunately, the function $M_{1}\left( r\right) $ is rather complicated
(it can be expressed in terms of the polylogarithms) and, once again, to avoid
unnecessary proliferation of long formulas it will not be presented here.
Interested reader is referred to~\cite{ja_grg}.

Generally, for the nonexterme black hole one has 
\begin{equation}
\mathcal{S}=\pi r_{0}^{2}+2\pi r_{1}\varepsilon +32\pi \varepsilon r_{0}^{2}
\left[ \frac{4\alpha }{r_{0}^{2}}M_{0}^{\prime }\left( r_{0}\right) +\frac{
2\alpha +\beta }{r_{0}}M_{0}^{\prime \prime }\left( r_{0}\right) \right] +
\mathcal{O}\left( \varepsilon ^{2}\right) .  \label{S_exp}
\end{equation}
On the other hand, making use of (\ref{REH}), the equation (\ref{S_exp}) can
be rewritten in the equivalent form 
\begin{equation}
\mathcal{S}=\pi r_{0}^{2}+ \varepsilon \frac{4\pi M_{1}\left( r_{0}\right) }{
1-2M_{0}^{\prime }\left( r_{0}\right) }+32\pi \varepsilon r_{0}^{2}\left[ 
\frac{4\alpha }{r_{0}^{2}}M_{0}^{\prime }\left( r_{0}\right) +\frac{2\alpha
+\beta }{r_{0}}M_{0}^{\prime \prime }\left( r_{0}\right) \right] +\mathcal{O}
\left( \varepsilon ^{2}\right) .  \label{S_second}
\end{equation}

The extremal case should be analyzed separately. The extremal configuration
of the ABGB black hole being the solution of the Einstein gravity is
described by 
\begin{equation}
\left| Q_{c}\right| =2w^{1/2}\mathcal{M}
\end{equation}
and 
\begin{equation}
r_{c}=\frac{4w}{1+w}\mathcal{M.}
\end{equation}
One expects, that the higher-order curvature terms modify these relations,
shifting (in a space of the parameters) extremal solution into a slightly
different position. Indeed, treating $M_{0}$ as a function of $Q^{2}$ and $
r,$ after some algebra, one concludes that the extremal configuration is
still possible and is described by the relations 
\begin{equation}
Q^{2}=Q_{c}^{2}+\varepsilon \Delta ,\,\,\,\,r_{+}=r_{c}+\varepsilon \delta ,
\label{Q_big}
\end{equation}
where 
\begin{equation}
\Delta =-\left( \dfrac{\partial }{\partial Q^{2}}M_{0}\right) ^{-1}M_{1}.
\label{del_big}
\end{equation}
and 
\begin{equation}
\delta =-\left( \dfrac{\partial ^{2}}{\partial r^{2}}M_{0}\right)
^{-1}\left( \dfrac{\partial }{\partial r}M_{1}\right) +\left( \dfrac{
\partial }{\partial Q^{2}}M_{0}\dfrac{\partial ^{2}}{\partial r^{2}}
M_{0}\right) ^{-1}\left( M_{1}\dfrac{\partial ^{2}}{\partial r\partial Q^{2}}
M_{0}\right) .
\end{equation}
Both $\delta $ and $\Delta $ are to be calculated for the parameters
describing extremal zeroth-order solution. Numerically, one has 
\begin{equation}
\Delta =\frac{1.05314}{\mathcal{M}}\alpha +\frac{0.43288}{\mathcal{M}}\beta
\end{equation}
and 
\begin{equation}
\delta =-\frac{0.05121}{\mathcal{M}}\alpha -\frac{0.57553}{\mathcal{M}}\beta
.
\end{equation}
Since the calculations of the entropy follow the general scheme sketched in
previous section they will not be presented here.

The purpose of the present paper (besides importance of the quadratic
gravity in its own and the natural curiosity) is twofold. First, one can
treat the calculations presented in this paper as the first step in
understanding of the influence of the higher curvature terms on the entropy
of black holes in a more complex setting than Maxwell electrodynamics. The
next step would involve, for example, inclusion of the all curvature
invariants of the order 4 and 6 and degree 2 and 3 \cite{wybourne,Lu,MTT}.
Moreover, it would be interesting to extend this analysis to general
D-dimensional manifolds. The natural candidate for a higher-curvature theory
is the Lovelock gravity~\cite{Lovelock}. Moreover, one may consider the more
general curvature terms, with arbitrary coefficients rather than those
inspired by particular theory.(See, for example~\cite{Deser2005,wybourne}
and references cited therein.) On the other hand, and this is even more
interesting, one can regard this sort of calculations as the preliminary
results allowing to analyse and understand the typical subtleties one is
likely to encounter when studying the semi-classical equations with the
source term given by the renormalized stress-energy tensor of the quantized
massive fields. Of course, the semi-classical equations are extremely
complex~\cite{JaD61,Kocio1}, but the general pattern that lies behind the 
calculations should
remain the same. This group of problems are currently actively investigated
and the results will be published elsewhere.


\end{document}